\documentclass[twocolumn,showpacs,preprintnumbers,amsmath,amssymb]{revtex4}
\usepackage[dvips]{graphicx}
\usepackage{dcolumn}
\usepackage{color}
\usepackage{amsmath}

\def\cor#1{{#1}}
\def\ket#1{  \left\vert  #1   \right\rangle   }

\def\etal{\textit{et al.}}

\begin{document}

\title{Entanglement creation with negative index metamaterials}

\author{Michael Siomau$^{1}$, Ali A. Kamli$^{1}$, Sergey A. Moiseev$^{2}$ and Barry C. Sanders$^{3}$}

\affiliation{$^{1}$Physics Department, Jazan University, P.O.~Box
114, 45142 Jazan, Kingdom of Saudi Arabia}

\affiliation{$^{2}$Kazan Physical-Technical Institute of Russian
Academy of Science, 420029 Kazan, Russia}

\affiliation{$^{3}$Institute for Quantum Information Science,
University of Calgary, Alberta T2N 1N4, Canada}

\date{\today}

\begin{abstract}
We propose a scheme for creating of a maximally entangled state
comprising two field quanta. In our scheme, two weak light fields,
which are initially prepared in either coherent or polarization
states, interact with a composite medium near an interface between a
dielectric and a negative index metamaterial. Such interaction leads
to a large Kerr nonlinearity, reduction of the group velocity of the
light and significant confinement of the light fields while
simultaneously avoiding amplitude losses of the incoming radiation.
All these considerations make our scheme efficient.
\end{abstract}

\pacs{03.67.Bg, 42.50.Dv, 42.50.Gy, 03.67.Hk}

\maketitle

Entanglement is an essential feature of quantum theory, which
manifests impressive advantages of recently established quantum
technologies over their classical counterpart
\cite{Horodecki:09,Guehne:09}. Variety of physical systems, starting
from atoms, electrons and photons \cite{Amico:08} and ending with
sophisticated molecules and even living organisms \cite{Romero:10},
can exhibit quantum entanglement. Among these systems, photons, the
quanta of electromagnetic field, take privileged position because of
their exceptional properties. Photons can be easily generated and
measured and can carry information over long distances being
resistant to detrimental effect of decoherence \cite{Braunstein:05}.
At the same time, photons do not directly interact with each other,
what makes difficult to prepare them in entangled states.

A conventional way to create entanglement of light quanta is through
their interaction with nonlinear (Kerr) medium, which have
intensity-dependent refractive index \cite{Sanders:11}. In natural
medium, however, Kerr nonlinearity is very small. Therefore, to
achieve significant entanglement of photons, one has to increase
both intensity of the field pulses and interaction time with the
medium. Such actions may not be always possible in practice, because
of diffraction and finite size of the medium, and are very unlikely
from the viewpoint of applications \cite{Braunstein:05,Kok:07},
where weak light fields (i.e. of energy of a single photon) are
desired.

In 2000, a pioneering proposal to ``design'' material exhibiting
strong nonlinear interaction at the single-photon level was made by
Lukin and Imamo\u{g}lu \cite{Lukin:00}. This idea stimulated a
number of theoretical investigations
\cite{Petrosyan:02,Wang:06,Yavuz:10} as well as experiments
\cite{Kang:03,Chen:06,Li:08}, to name just a few. However, despite
these remarkable results in achieving large Kerr nonlinearity,
efficient creation of entanglement at the low energy limit remains
challenging in many aspects \cite{Shapiro:06,Gea:10}.

\begin{figure}[b]
\begin{center}
\includegraphics[scale=0.53]{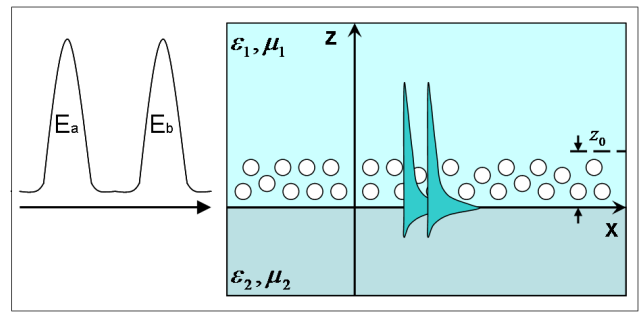}
\caption{(Color online) Two weak light pulses create two surface
polaritons near the interface between the dielectric $z>0$ and the
metamaterial $z<0$. While electromagnetically induced transparency
is established for both polaritons simultaneously, they will
propagate along the interface with small group velocities
$\upsilon_g < c$ and interact nonlinearly with each other.}
 \label{fig-1}
\end{center}
\end{figure}

In this work we suggest a scheme in which initially uncorrelated
states of light field become entangled due to their interaction with
a medium near an interface between a dielectric and a negative index
metamaterial. The medium of interest consists of a dielectric (which
has a layer of thickness $z_0$ doped with six level atoms
\cite{Petrosyan:02}) and a metamaterial placed together, as shown in
Fig.~\ref{fig-1}. Due to interaction with the medium, an incident
light beam creates a spatially confined surface polariton
\cite{Maier:07} which propagates along the interface with
substantially reduced group velocity $\upsilon_g < c$. Although in
natural mediums a surface polariton undergos slashing amplitude
loss, specific design of the medium makes possible to suppress
losses significantly in a narrow frequency bandwidth of the incoming
light \cite{Kamli:08}. Placing the layer of six level atoms near the
interface allows us to establish double electromagnetically induced
transparency \cite{Fleisch:05} (i.e. for two incoming pulses
simultaneously) and, at the same time, create large Kerr
nonlinearity \cite{Petrosyan:02}. All mentioned factors contribute
to an efficient nonlinear interaction between the two surface
polaritons in the medium. Such interaction makes possible a mutual
$\pi$-phase shift between the polaritons which leads to entanglement
of the light fields.

Ignoring presence of the atomic layer near the interface, the
process of interaction between the light fields and the medium can
be considered from the viewpoint of classical electrodynamics.
Macroscopic properties of the material can be characterized with
electric permittivity $\varepsilon$ and magnetic permeability $\mu$.
For a dielectric, these parameters are strictly positive, while both
of them may be simultaneously negative for a metamaterial
\cite{Smith:04}. Let us assume that the dielectric has constant
homogeneous parameters $\varepsilon_1$ and $\mu_1$, while parameters
$\varepsilon_2$ and $\mu_2$ are frequency-dependent for the
metamaterial and are given by \cite{Kamli:08}
\begin{eqnarray}
 \label{dispersion-relations}
\varepsilon_2 (\omega) &=& \varepsilon_\infty -
\frac{\omega_e^2}{\omega \, (\omega + i\, \gamma_e)} \, , \nonumber
 \\[0.1cm]
\mu_2 (\omega) &=& \mu_\infty - \frac{\omega_m^2}{\omega \, (\omega
+ i\, \gamma_m)} \, ,
\end{eqnarray}
where $\omega_e$ and $\omega_m$ are electric and magnetic plasma
frequencies, $\gamma_e$ and $\gamma_m$ are corresponding (empiric)
decay rates and $\varepsilon_\infty$ and $\mu_\infty$ are background
constants \cite{Maier:07}. \cor{Here we have chosen the simplest
(Drude-like) model for magnetic permeability $\mu_2 (\omega)$. This
model is known to be adequate in the optical region
\cite{Maier:07,Merlin:09}, although more sophisticated models can be
taken into consideration \cite{Kamli:10}.}

Electromagnetic field of the surface polaritons can be found form
Maxwell equations with boundary conditions for $\varepsilon_i$ and
$\mu_i$ $(i=1,2)$. Since the permittivity and the permeability
(\ref{dispersion-relations}) may be simultaneously negative, both
transverse magnetic (TM) and transverse electric (TE) polarizations
of the electromagnetic field may exist in the medium. Natural
mediums, in contrast, support only TM polarization \cite{Maier:07}.
To be specific, we shall later focus on the TM waves. The wave
vector of the electromagnetic field in the medium is given by the
dispersion relation as follows
\begin{equation}
 \label{dispersion-k}
K(\omega) \, = \, \frac{\omega}{c} \: \sqrt{\varepsilon_1 \,
\varepsilon_2 \, \frac{\varepsilon_2 \, \mu_1 - \varepsilon_1 \,
\mu_2}{\varepsilon_2^2 - \varepsilon_1^2}} \, .
\end{equation}
The real part of this expression gives dispersion of the field,
while its imaginary part stands for absorbtion loss.

It has been shown that absorbtion loss can be completely suppressed
in a narrow frequency bandwidth due to destructive interference of
electric and magnetic absorbtion responses of the medium
\cite{Kamli:08}. For example, taking $\varepsilon_1 = 1.3$ and
$\mu_1 = 1$ and $\omega_e = 1.37 \times 10^{16} {\rm s}^{-1}$,
$\gamma_e = 2.73 \times 10^{13} {\rm s}^{-1}$ (as for Ag) and
assuming $\omega_m = 10^{15} {\rm s}^{-1}$, $\gamma_m = 10^{12} {\rm
s}^{-1}$, $\varepsilon_\infty = 5$ and $\mu_\infty = 5$, one can see
that absorbtion loss ${\rm Im}[K(\omega)]$ vanishes for $\omega_0
\approx 4.4 \times 10^{14} s^{-1} = 440$ THz, what corresponds to
red light of visible spectrum. It is important to note that
metamaterials with negative refractive index have been observed in
the red region of visible spectrum \cite{Smith:04}. More details
about the dispersion relation (\ref{dispersion-k}) and possible
parameters of the medium can be found in Ref.~\cite{Kamli:08} and
references therein.

We are now at the position to consider the interaction between the
surface polariton light fields and the medium quantum mechanically.
Electric field of each of the surface polaritons can be quantized
near the surface and written in the plane wave expansion as
\cite{Scully:97}
\begin{equation}
 \label{quantization}
\textbf{E}(\textbf{r},t)=\int {\rm d} k \left[ {\bf E}_0(k)\, a(k)\,
e^{i k x -\omega t} + {\rm H.c.} \right] \, .
\end{equation}
Here we introduced $k \equiv {\rm Re}[K(\omega)]$, taking into
account that the wave vector is approximated by its real part
$k(\omega)\approx {\rm Re}[K(\omega)]$ in the low loss frequency
range. The amplitude ${\bf E}_0(k)$ can be found from the
requirement that it should obey the field Hamiltonian ${\rm H}_F =
1/2 \int d^3 r \, [\tilde{\varepsilon} \, <|{\bf E}|^2> +
\tilde{\mu} <|{\bf H}|^2>]$ in a dispersive lossless medium
\cor{\cite{Kamli:10}. It is important to note that our quantization
procedure is applicable only in a narrow frequency bandwidth where
the losses are low. In general case, the quantization of
electromagnetic field in dispersive and absorptive media is a much
more complicated task \cite{Bhat:06,Chen:10,Ginzburg:12}.}

\begin{figure}
\begin{center}
\includegraphics[scale=0.6]{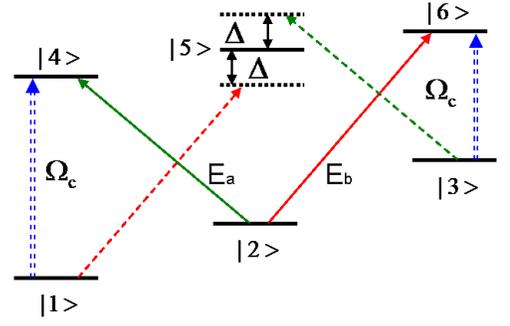}
\caption{(Color online) Configuration of a six level atom for
creating double electromagnetically induced transparency for two
weak fields $E_a$ and $E_b$. Here, the fields $E_a$ and $E_b$ are
coupled with transitions $\ket{2} \rightarrow \ket{4}$ and $\ket{2}
\rightarrow \ket{6}$ resonantly and with transitions $\ket{3}
\rightarrow \ket{5}$ and $\ket{1} \rightarrow \ket{5}$ off
resonantly with detuning $\Delta$. Two classical control fields
drive transitions $\ket{1} \rightarrow \ket{4}$ and $\ket{3}
\rightarrow \ket{6}$ \cite{Petrosyan:02}.}
 \label{fig-2}
\end{center}
\end{figure}

The quantized polariton fields exhibit a remarkable property of
confinement along $z$ direction. The confinement can be quantified
as $\xi(\omega) = 1/{\rm Re} [k^\perp(\omega)]$, where $k^\perp
(\omega) = \sqrt{K^2(\omega) - \omega^2 \varepsilon_1 \, \mu_1/c^2}$
is the normal component of the real part of the wave vector
(\ref{dispersion-k}). This property of the quantized fields ensure
interaction of the polaritons with the six level atoms embedded into
the dielectric and defines effective thickness of the atomic layer
$\xi(\omega) \approx z_0$ in practice. The reason of injection the
six level atoms near the dielectric-metamaterial interface, is that
such atomic system has been shown to cause the effect of double
electromagnetically induced transparency simultaneously exhibiting
symmetric nonlinearity for two incoming pulses \cite{Petrosyan:02}.
The energy levels of the system are shown schematically in
Fig.~\ref{fig-2}. The model of a six level atom can be practically
realized on the rubidium isotope ${}^{87}{\rm Rb}$
\cor{\cite{Petrosyan:02}}, for example.

The interaction of the surface polaritons and the six level atoms
can be modeled with electric dipole hamiltonian ${\rm H}_{ED} =
-\sum {\bf d}_i \cdot {\bf E} ({\bf r}_i)$, where ${\bf E} ({\bf
r}_i)$ is the electric field (\ref{quantization}) of the surface
polaritons, ${\bf r}_i$ the position of the atoms and the summation
is to be done over all atoms in the interaction volume
\cite{Kamli:10}.

\begin{figure}
\begin{center}
\includegraphics[scale=0.9]{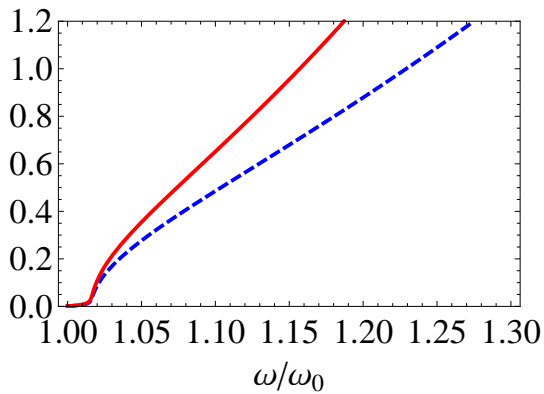}
\caption{(Color online) Kerr coefficient $\chi \times 10^4$ (dashed
blue) and corresponding mutual phase shift $\phi$ in units $\pi$
(solid red) as functions of the field frequency $\omega/\omega_0$.}
 \label{fig-3}
\end{center}
\end{figure}

Because of the symmetry of the atomic levels structure the two
surface polaritons propagate in the medium with equal group
velocities $\upsilon_a = \upsilon_b \equiv \upsilon_g$
\cite{Petrosyan:02}. Dynamics of the surface polariton field
operators can be obtained in Heisenberg picture by solving
corresponding set of equations
\begin{equation}
 \label{heisenberg-eq}
\left( \frac{1}{\upsilon_g} \, \frac{\partial}{\partial t} +
\frac{\partial}{\partial x} \right) \, \textbf{E}_n(\textbf{r},t) =
i \, \chi \, I_m \, \textbf{E}_n(\textbf{r},t) \, ,
\end{equation}
where the adiabatic approximation has been used to ignore time
derivatives of higher order \cite{Kamli:10}. Here $n,m=a,b \, (n\neq
m)$, $I_m = |\textbf{E}_m(\textbf{r},t)|^2$ and $\chi$ is Kerr
coefficient given by
\begin{equation}
 \label{kerr-nonlin}
\chi = \frac{2\pi n z_0 \, f[(k_a+k_b-k_c)z_0]}{\hbar^4 \upsilon_g
|\Omega_c|^2 \, \Delta} \, <|{\bf d}_{24}{\bf E}_a|^2|{\bf
d}_{26}{\bf E}_b|^2> \, ,
\end{equation}
where $n$ is atomic density, $z_0$ is thickness of the atomic layer,
$f[x]\equiv (e^{-x} \sinh{x})/x$, $k_a$ and $k_b$ are the real parts
of the polaritons wave numbers, $k_c$ and $\Omega_c$ are the wave
number and the Rabi frequency of the driving field, $\upsilon_g$ is
group velocity of the polaritons ignoring the atomic layer, $\Delta$
stands for spectral detuning, ${\bf d}_{24}$ and ${\bf d}_{26}$ give
atomic dipole moments of the corresponding transitions, ${\bf E}_a$
and ${\bf E}_b$ are electric field operators of the polaritons and
$<...>$ denotes averaging over orientation of the dipole moments.
Typical atomic density in a gas is $2 \times 10^{14} {\rm cm}^{-3}$.
To establish double electromagnetically induced transparency in
${}^{87}{\rm Rb}$, Rabi frequency of the control field is to be
$\Omega_c = 1 {\rm MHz}$, the transition wavelength is $780 {\rm
nm}$, the detuning is $\Delta=1.4 {\rm MHz}$ and the dipole moments
are about $5 e a_0$, where $e$ is the electron charge $a_0$ is the
Bohr radius. Assuming the thickness of the atomic layer $z_0 = 2
\mu{\rm m}$, we obtain Kerr nonlinearity as displayed in
Fig.~\ref{fig-3}.

Although the Kerr nonlinearity $\chi$ is of order of $10^{-4}$, a
significant mutual phase shift of order of unity can be achieved
between the surface polaritons. The mutual phase shift is given by
$\phi = \chi \, \omega \, L \, / \upsilon_g $, where $\omega$ is the
light frequency, $L$ is the length of interaction in the medium and
$\upsilon_g$ is the group velocity of the light in the medium
ignoring the layer of the five level atoms. For chosen parameters of
the medium $\upsilon_g \approx 0.4 c$ and assuming $L = 1 {\rm mm}$,
the mutual phase shift is shown in Fig.~3. The surface polaritons
receive a mutual phase shift of order of $\pi$ at frequency
$\omega_\pi \approx 1.24 \, \omega_0 = 545 \, {\rm THz}$ (green
light) which is close to the no-loss frequency $\omega_0$.

Here we would like to point out that, because of the symmetry of the
levels of the six level atom, the refractive index of the medium is
exactly the same for two surface polaritons. That is why the
polaritons propagate in the medium with equal group velocities and
experience identical nonlinearity. Alternatively, five level atoms
\cite{Wang:06} can be placed near dielectric-metamaterial interface
\cite{Moiseev:10}. In this case, two surface polaritons propagate in
the medium with different group velocities and experience different
nonlinearity in the double electromagnetically induced transparency
regime. Latter scheme is best suitable to achieve a uniform
cross-phase modulation \cite{Marzlin:10}.

The mutual (symmetric) phase shift can be used to create
entanglement between initially uncorrelated field modes. For
single-mode incident fields, the interaction of the surface
polaritons in the (Kerr) medium can be described with the help of an
effective Hamiltonian
\begin{equation}
 \label{effect-ham}
{\rm H}_{\rm eff} \, = \, \hbar \, \chi \, a^\dag a \, b^\dag b \, ,
\end{equation}
where $a^\dag, \, b^\dag$ and $a, \, b$ are creation and
annihilation operators of the field modes of the two polaritons.
Time evolution of these operators is initiated by the unitary
transformation $U(t)= \exp (-i\, \phi \, a^\dag a \, b^\dag b)$ and
is given by
\begin{equation}
 \label{unit-transf}
a(t) = e^{-i \phi \, b^\dag b} a(0) \, , \hspace{0.4cm} b(t) = e^{-i
\phi \, a^\dag a} b(0) \, .
\end{equation}

If the initial states of the incident fields are uncorrelated
single-mode coherent states $\ket{\alpha}$ and $\ket{\beta}$
\cite{Scully:97}, the final state $\ket{\psi(t)}_{ab}$ after the
interaction can be written in the Fock basis as
\cite{Paternostro:03}
\begin{equation}
 \label{coherent}
\ket{\psi(t)}_{ab} = e^{-\frac{|\beta|^2}{2}} \sum_{n=0}^\infty
\frac{\beta^n}{\sqrt{n!}} \, \ket{\alpha\, e^{-i \phi n}}_a \otimes
\ket{n}_b \, ,
\end{equation}
where the dynamics of the creation and annihilation operators
(\ref{unit-transf}) has been taken into account. Assuming $\phi =
\pi$ and decomposing the sum in Eq.~(\ref{coherent}) over odd and
even values of the index $n$, we obtain the following form of the
final state
\begin{equation}
\ket{\psi}^f_{ab} = \frac{1}{2} \left( \ket{\alpha}_a \left(
\ket{\beta}_b + \ket{-\beta}_b \right) + \ket{-\alpha}_a \left(
\ket{\beta}_b - \ket{-\beta}_b \right) \right) \, . \nonumber
\end{equation}
This state is local unitary equivalent to the entangled state
$\left( \ket{\alpha}_a\ket{\beta}_b + \ket{-\alpha}_a\ket{-\beta}_b
\right)/\sqrt{N}$ \cite{Paternostro:03}, where
$N=2-2\exp(-2|\alpha|^2-2|\beta|^2)$, which is known to preserve
exactly one e(ntangled )bit of quantum information \cite{Enk:01}.

In contrast to our assumption above, the initial states
$\ket{\psi}_a$ and $\ket{\psi}_b$ of the incident field can be also
assumed to be polarization states of photons. In this case,
entanglement of the field modes can be achieved, for example, with
the help of Nemoto-Munro protocol \cite{Nemoto:04}. To understand
this protocol better, let us assume that the incident field $a$ is
prepared into a superposition of vacuum and a single photon states
and is given by $\ket{\psi}_a = c_0 \ket{0} + c_1 \ket{1}$ in Fock
basis. The second incident field is prepared into a coherent state
$\ket{\alpha}_b$. When the two fields interact with the media, the
resulting state of the fields is given by
\begin{equation}
U(t)\, \ket{\psi}_a \ket{\alpha}_b = c_0 \ket{0} \ket{\alpha}_b +
c_1 \ket{1} \ket{\alpha \, e^{i \phi}}_b \, .
\end{equation}
The state of the field $a$ is unaffected by the interaction, while
the state of the field $b$ receives a phase shift, which is
proportional to the number of photons on the the state
$\ket{\psi}_a$.

Assume, we have two polarization qubits to become entangled. The
qubits are initially prepared in single-photon superposition states
$\ket{\psi}_a = c_0 \ket{H} + c_1 \ket{V}$ and $\ket{\psi}_b = d_0
\ket{H} + d_1 \ket{H}$, where $\ket{V}$ and $\ket{H}$ are
polarization degrees of freedom. These qubits are split individually
on polarizing beam splitters into spatial modes and interact with an
additional probe beam (which is in a coherent state
$\ket{\alpha}_p$) in the Kerr medium. The resulting state of the
three beams is given by
\begin{eqnarray}
\ket{\psi}_{abc} & = & \left( c_0 d_0 \ket{HH} + c_1 d_1
\ket{VV}\right) \ket{\alpha}_p \nonumber
 \\[0.1cm]
& & \hspace{-1.1cm} + \, c_0 d_1 \ket{HV}\ket{\alpha e^{i \phi}}_p +
c_1 d_0 \ket{VH}\ket{\alpha e^{-i \phi}}.
\end{eqnarray}
The first term in this expression does not receive any phase shift,
while the the second and the third terms receive opposite sign
shifts. This makes possible to transform the three-party state into
entangled (Bell) bipartite states by performing a homodyne
measurement on the probe. The measurement results into either $c_0
d_0 \ket{HH} + c_1 d_1 \ket{VV}$ or $c_0 d_1 \ket{HV} + c_1 d_0
\ket{VH}$ states, which are  both maximally entangled states of
qubits for $c_0=c_1=d_0=d_1=1/\sqrt{2}$. It is also important to
note that the described above Nemoto-Munro protocol allows us to
construct entangling Controlled-NOT gate \cite{Kok:07} with large
Kerr nonlinearity, opening prominent possibility to use
metamaterials in quantum computing.

Presented scheme for entanglement creation with negative index
metamaterials may also find its applications in quantum
communication and quantum teleportation with both coherent
\cite{Braunstein:05,Enk:01} and polarization states \cite{Kok:07}.
Moreover, Kerr nonlinearity, created with the described medium, can
be used to generate multimode entangled coherent states
\cite{Enk:03} and multiphoton Greenberger-Horne-Zeilinger states
\cite{Jin:07}.

We also would like to outline that in the present discussion we
restricted ourselves with TM polarization of surface polaritons. As
we mentioned before, both TM and TE polarizations may exist on the
dielectric-metamaterial interface. These polarizations may be used
for information encoding on a par with encoding in quantum states of
the field quanta. Another attractive idea is to use a trade-off
between confinement and losses of the surface polaritons
\cor{\cite{Kamli:10,Moiseev:10}}. This trade-off may be used to
establish two regimes, corresponding to ``manipulation'' and
``low-loss transmission'', which are highly desired in quantum
computation \cite{Ladd:10}. Both mentioned possibilities will be the
subject of further investigations.

In conclusion, we presented the scheme for entanglement creation
with the composite medium consisting of the dielectric and the
negative index metamaterial. The surface polaritons, which are
created by the incident light in the medium, propagate along the
dielectric-metamaterial interface with substantially reduced group
velocity, exhibiting property of spatial confinement and with
suppressed amplitude losses. Placing a layer of six level atoms near
the interface allowed us to establish symmetric nonlinear
interaction between the surface polaritons, which can be utilized to
create entanglement between initially uncorrelated coherent or
polarization states of light.


\begin{thebibliography}{30}

\bibitem{Horodecki:09}
   R.~Horodecki \etal, Rev.\ Mod.\ Phys. \textbf{81}, 865 (2009).

\bibitem{Guehne:09}
   O.~G\"{u}hne and G.~T\'{o}th, Phys. Rep. \textbf{474}, 1 (2009).

\bibitem{Amico:08}
   L.~Amico \etal, Rev. Mod. Phys. \textbf{80}, 517 (2008).

\bibitem{Romero:10}
   O.~Romero-Isart \etal, New J. Phys. \textbf{12}, 033015 (2010).

\bibitem{Braunstein:05}
   S.L.~Braunstein and P.~van Loock, Rev. Mod. Phys. \textbf{77},
   513 (2005).

\bibitem{Sanders:11}
   B.C.~Sanders, arXiv:1112.1778 (2011).

\bibitem{Kok:07}
   P.~Kok \etal, Rev. Mod. Phys. \textbf{79}, 135 (2007).

\bibitem{Lukin:00}
   M.D.~Lukin and A.~Imamo\u{g}lu, Phys. Rev. Lett. \textbf{84}, 1419
   (2000).

\bibitem{Petrosyan:02}
   D.~Petrosyan and G.~Kurizki, Phys. Rev. A \textbf{65}, 033833
   (2002).

\bibitem{Wang:06}
   Z.-B.~Wang, K.-P.~Marzlin and B.C.~Sanders, Phys. Rev. Lett.
   \textbf{97}, 063901 (2006).

\bibitem{Yavuz:10}
   D.D.~Yavuz and D.E.~Sikes, Phys. Rev. A \textbf{81}, 035804
   (2010).

\bibitem{Kang:03}
   H.~Kang and Y.~Zhu, Phys. Rev. Lett. \textbf{91}, 093601 (2003).

\bibitem{Chen:06}
   Y.-F.~Chen \etal, Phys. Rev. Lett. \textbf{96}, 043603 (2006).

\bibitem{Li:08}
   S.~Li \etal, Phys. Rev. Lett. \textbf{101}, 073602 (2008).

\bibitem{Shapiro:06}
   J.H.~Shapiro, Phys. Rev. A \textbf{73}, 062305 (2006).

\bibitem{Gea:10}
   J.~Gea-Banacloche, Phys. Rev. A {\bf 81}, 043823 (2010).

\bibitem{Maier:07}
   S.A.~Maier, \textit{Plasmonics: Fundamentals and Applications}
   (Springer, 2007).

\bibitem{Merlin:09}
   R.~Merlin, Proc. Nat. Acad. Sci. \textbf{106}, 1693 (2009).

\bibitem{Kamli:08}
   A.A.~Kamli, S.A.~Moiseev and B.C.~Sanders, Phys. Rev. Lett.
   \textbf{101}, 263601 (2008).

\bibitem{Fleisch:05}
   M.~Fleischhauer, A.~Imamo\u{g}lu and J.P.~Marangos, Rev. Mod.
   Phys. \textbf{77}, 633 (2005).

\bibitem{Smith:04}
   V.M.~Shalaev, Nature Photonics \textbf{1}, 41 (2007).

\bibitem{Kamli:10}
   A.A.~Kamli, S.A.~Moiseev and B.C.~Sanders, Int. J. Quant. Inf.
   \textbf{9}, 263 (2011).

\bibitem{Scully:97}
   M.O.~Scully and M.S.~Zubairy, \textit{Quantum Optics} (Cambridge
   University Press, 1997).

\bibitem{Bhat:06}
   N.A.R.~Bhat and J.E.~Sipe, Phys. Rev. A {\bf 73}, 063808 (2006).

\bibitem{Chen:10}
   P.Y.~Chen \etal, Phys. Rev. A {\bf 82}, 053825 (2010).

\bibitem{Ginzburg:12}
   P.~Ginzburg and A.V.~Zayats, Opt. Expr. {\bf 20}, 6720 (2012).

\bibitem{Moiseev:10}
   S.A.~Moiseev, A.A.~Kamli and B.C.~Sanders, Phys. Rev. A
   \textbf{81}, 033839 (2010).

\bibitem{Marzlin:10}
   K.-P.~Marzlin \etal, J. Opt. Soc. Am. B {\bf 27}, {\rm A36} (2010).

\bibitem{Paternostro:03}
   M.~Paternostro, M.S.~Kim and B.S.~Ham, Phys. Rev. A \textbf{67}, 023811 (2003).

\bibitem{Enk:01}
   S.J.~van Enk and O.~Hirota, Phys. Rev. A \textbf{64}, 022313
   (2001).

\bibitem{Nemoto:04}
   K.~Nemoto and W.J.~Munro, Phys. Rev. Lett. \textbf{93}, 250502
   (2004).

\bibitem{Enk:03}
   S.J.~van Enk, Phys. Rev. Lett. \textbf{91}, 017902 (2003).

\bibitem{Jin:07}
   G.-S.~Jin, Y.~Lin and B.~Wu, Phys. Rev. A \textbf{75}, 054302 (2007).

\bibitem{Ladd:10}
   T.D.~Ladd \etal, Nature \textbf{464}, 45 (2010).

\end{thebibliography}
\end{document}